\documentclass[twocolumn,pre,aps,nofootinbib,superscriptaddress,showpacs,preprintnumbers]{revtex4}

\usepackage{graphicx}
\usepackage{bm}
\graphicspath{{./fig/}{./png/}}
\usepackage{amsmath}

\newcommand{\EQ}{\begin{equation}}
\newcommand{\EN}{\end{equation}}
\newcommand{\EQA}{\begin{eqnarray}}
\newcommand{\ENA}{\end{eqnarray}}
\newcommand{\eq}[1]{(\ref{#1})}
\newcommand{\EEq}[1]{Equation~(\ref{#1})}
\newcommand{\Eq}[1]{Eq.~(\ref{#1})}

\newcommand{\Fig}[1]{Figure~\ref{#1}}

{}
{}
\newcommand{\meanFFFF}{\overline{\mbox{\boldmath ${\cal F}$}}{}}{}
\newcommand{\meanemf}{\overline{\cal E} {}}

{}
{}
\newcommand{\meanEMF}{\overline{\mbox{\boldmath ${\cal E}$}}{}}{}
\newcommand{\meanEEEE}{\overline{\mbox{\boldmath ${\cal E}$}}{}}{}
{}
{}
{}
{}
{}
\newcommand{\meanBB}{\overline{\mbox{\boldmath $B$}}{}}{}
{}
{}
{}
{}
{}
{}
{}
{}
{}
{}
\newcommand{\meanUU}{\overline{\bm{U}}}

{}
{}
{}

\newcommand{\meanB}{\overline{B}}

\newcommand{\meanC}{\overline{C}}

\newcommand{\meanFFF}{\overline{\cal F}}

{}

{}
{}

\newcommand{\kk}{\bm{k}}

\newcommand{\xx}{\bm{x}}
\newcommand{\XX}{\bm{X}}

\newcommand{\uu}{\mbox{\boldmath $u$} {}}
\newcommand{\UU}{\mbox{\boldmath $U$} {}}

\def\bb{\bm{b}}

\newcommand{\BB}{\mbox{\boldmath $B$} {}}

\newcommand{\GG}{\mbox{\boldmath $G$} {}}

\newcommand{\nab}{\mbox{\boldmath $\nabla$} {}}
\newcommand{\cchi}{\bm{\chi}}
\newcommand{\ppsi}{\mbox{\boldmath $\psi$} {}}
\newcommand{\xxi}{\mbox{\boldmath $\xi$} {}}
\newcommand{\ii}{{\rm i}}

\newcommand{\dd}{{\rm d} {}}

\def\ga{\mathrel{\mathchoice {\vcenter{\offinterlineskip\halign{\hfil
$\displaystyle##$\hfil\cr>\cr\sim\cr}}}
{\vcenter{\offinterlineskip\halign{\hfil$\textstyle##$\hfil\cr>\cr\sim\cr}}}
{\vcenter{\offinterlineskip\halign{\hfil$\scriptstyle##$\hfil\cr>\cr\sim\cr}}}
{\vcenter{\offinterlineskip\halign{\hfil$\scriptscriptstyle##$\hfil\cr>\cr\sim\cr}}}}}
\def\St{\mbox{\rm St}}

\def\Rm{\mbox{Rm}}

\def\Pe{\mbox{\rm Pe}}

\def\kf{k_{\rm f}}

\def\urms{u_{\rm rms}}

\def\qeta{q_\eta}
\def\qkap{q_\kappa}

\def\kappat{\kappa_{\rm t}}

\def\kappatz{\kappa_{\rm t0}}

\def\etat{\eta_{\rm t}}
\def\etatz{\eta_{\rm t0}}

\def\half{{\textstyle{1\over2}}}

\def\onethird{{\textstyle{1\over3}}}

\newcommand{\yana}[3]{, Astron. Astrophys. {\bf #2}, #3 (#1).}
\newcommand{\pana}[1]{, Astron. Astrophys., in press (#1).}

\newcommand{\ymn}[3]{, Mon.\ Not.\ R.\ Astron.\ Soc.\ {\bf #2}, #3 (#1).}

\newcommand{\yjfm}[3]{, J. Fluid Mech. {\bf #2}, #3 (#1).}

\newcommand{\yprd}[3]{, Phys.\ Rev.\ D {\bf #2}, #3 (#1).}
\newcommand{\ypre}[3]{, Phys.\ Rev.\ E {\bf #2}, #3 (#1).}
\newcommand{\yprl}[3]{, Phys.\ Rev.\ Lett.\ {\bf #2}, #3 (#1).}
\newcommand{\sprl}[1]{, Phys.\ Rev.\ Lett., submitted (#1).}

\newcommand{\yapj}[3]{, Astrophys. J. {\bf #2}, #3 (#1).}

\newcommand{\yapjl}[3]{, Astrophys. J. {\bf #2}, #3 (#1).}

\newcommand{\ygafd}[3]{, Geophys. Astrophys. Fluid Dynam. {\bf #2}, #3 (#1).}

\newcommand{\yjour}[4]{, #2 {\bf #3}, #4 (#1).}

\newcommand{\yproc}[4]{, (ed. #4), pp. #2. #3 (#1).}
\newcommand{\ybook}[3]{, {\em #2}. #3 (#1).}

\begin{document}
\preprint{NORDITA 2011-32}

\title{Mean-field diffusivities in passive scalar and magnetic transport
in irrotational flows}

\author{Karl-Heinz R\"adler}
\email{khraedler@arcor.de}
\affiliation{Astrophysikalisches Institut Potsdam, An der Sternwarte 16,
D-14482 Potsdam, Germany}
\affiliation{NORDITA, AlbaNova University Center, Roslagstullsbacken 23,
SE-10691 Stockholm, Sweden}
\affiliation{Kavli Institute for Theoretical Physics,
University of California, Santa Barbara, CA 93106, USA}

\author{Axel Brandenburg}
\affiliation{NORDITA, AlbaNova University Center, Roslagstullsbacken 23,
SE-10691 Stockholm, Sweden}
\affiliation{Department of Astronomy, Stockholm University,
SE-10691 Stockholm, Sweden}
\affiliation{Kavli Institute for Theoretical Physics,
University of California, Santa Barbara, CA 93106, USA}

\author{Fabio Del Sordo}
\affiliation{NORDITA, AlbaNova University Center, Roslagstullsbacken 23,
SE-10691 Stockholm, Sweden}
\affiliation{Department of Astronomy, Stockholm University,
SE-10691 Stockholm, Sweden}

\author{Matthias Rheinhardt}
\affiliation{NORDITA, AlbaNova University Center, Roslagstullsbacken 23,
SE-10691 Stockholm, Sweden}

\date{\today,~ $ $Revision: 1.188 $ $}

\begin{abstract}
Certain aspects of the mean-field theory of turbulent passive scalar transport and of mean-field electrodynamics are considered
with particular emphasis on aspects of compressible fluids.
It is demonstrated that the total mean-field diffusivity for passive
scalar transport in a compressible flow may well be smaller than the molecular diffusivity.
This is in full analogy to an old finding regarding the magnetic mean-field diffusivity in an electrically conducting
turbulently moving compressible fluid.
These phenomena occur if the irrotational part of the motion dominates the vortical part,
the P\'eclet or magnetic Reynolds number is not too large, and, in addition, the variation of the flow pattern is slow.
For both the passive scalar and the magnetic cases several further analytical results on mean-field diffusivities
and related quantities found within the second-order correlation approximation are presented,
as well as numerical results obtained by the test-field method, which applies independently of this approximation.
Particular attention is paid to non-local and non-instantaneous connections between the turbulence-caused terms
and the mean fields.
Two examples of irrotational flows, in which interesting phenomena in the above sense occur, are investigated in detail.
In particular, it is demonstrated that the decay of a mean scalar in a compressible fluid under the influence of these flows
can be much slower than without any flow, and can be strongly influenced by the so-called memory effect,
that is, the fact that the relevant mean-field coefficients depend on the decay rates themselves.
\end{abstract}
\pacs{52.65.Kj, 52.30.Cv, 52.35.Vd }

\maketitle

\section{Introduction}

Many investigations of transport processes in turbulently moving fluids
have been done in the framework of the mean-field concept.
A simple example is the transport of a passive scalar quantity like the number density of particles in a turbulent fluid
\cite{EKR95,EKR96,EKR97,BSV09,HKRB11}.
Another important example is the magnetic-field transport in electrically conducting turbulent fluids.
The widely elaborated mean-field electrodynamics, or magnetofluiddynamics, delivers in particular the basis of
the mean-field theory of cosmic dynamos \cite{Mof78,KR80}.

The original equation governing the behavior of a passive scalar in a fluid contains a diffusion term
with a diffusion coefficient, say $\kappa$.
In the corresponding mean-field equation there appears,
in the simple case of isotropic turbulence,
the effective mean-field diffusivity $\kappa + \kappat$ in place of $\kappa$,
where $\kappat$ is determined by the turbulent motion
and therefore sometimes called ``turbulent diffusivity".
Likewise, the induction equation governing the magnetic field in an electrically conducting fluid contains a diffusion term
with the magnetic diffusivity $\eta$.
In the mean-field induction equation there appears, again for
isotropic turbulence, $\eta + \etat$ in place of $\eta$,
where $\etat$ is again determined by the turbulent
motion and, sometimes, called the ``turbulent magnetic diffusivity".

At first glance it seems plausible that turbulence enhances the
effective diffusion, corresponding to positive $\kappat$ and $\etat$.
In a compressible fluid, however, this is not always true.
A counterexample for the magnetic case has been long known.
Represent the velocity in the form $\uu = \nab \times \ppsi + \nab \phi$ by a vector potential $\ppsi$
satisfying
the gauge condition $\nab\cdot\ppsi=0$ and a scalar potential $\phi$.
Let $u_c$, $\lambda_c$, and $\tau_c$ be a characteristic magnitude, length,
and time, respectively, of the velocity field.
Assume that the magnetic Reynolds number $u_c \lambda_c / \eta$ is small compared to unity
and that $\tau_c$ considerably exceeds the free-decay time $\lambda_c^2 / \eta$ of a magnetic structure
of size $\lambda_c$.
Then it turns out \cite{KR80,Rae00,Rae07,RR07} that
\EQ
\etat = \frac{1}{3 \eta} (\overline{\ppsi^2} - \overline{\phi^2}) \,  .
\label{eq01}
\EN
That is, negative $\etat$ are well possible if the part of $\uu$
determined by the potential $\phi$ dominates.
Then, the mean-field diffusivity is smaller than the molecular one.
This surprising result
deserves more thorough examination,
which is indeed one of the motivations behind this paper.
Here it will be shown
that a result analogous to \eq{eq01} applies to $\kappa_{\rm t}$, too.
These results apply not only to turbulence in the narrow sense,
but also to other kinds of random and even non-random (including steady) flows.

Results of that kind might be of some interest for the turbulence
in the interstellar medium.
It is widely believed that it is mostly driven by supernova explosions
\cite{Korpi_etal99,dAML02,Bals04,Gre08}.
In this case the driving force, and so the flow, too,
could have noticeable irrotational parts, i.e., parts that are described by gradients of potentials.
However, when rotation or shear is important, or the Mach number is close
to or in excess of unity and the baroclinic effect present, vorticity
production becomes progressively more important---even when the forcing
of the flow is purely irrotational \cite{DSB11}.

Another possible application of such results could be in studies of the
very early Universe, where phase transition bubbles
are believed to be generated in connection with the electroweak phase transition
\cite{KKS86,Ingatius94}.
The relevant equation of state is that of an ultra-relativistic gas with
constant sound speed $c/\sqrt{3}$, where $c$ is the speed of light.
This is a barotropic equation of state, so the baroclinic term vanishes.
Hence, there is no obvious source of vorticity in the
(non-relativistic) bulk motion inside
these bubbles so that it should be essentially irrotational.
This changes, however, if there is a magnetic field of significant
strength, because the resulting Lorentz force is in general not a
potential one.

In Sec.~II of this paper we give an outline of the mean-field theory of passive scalar transport,
prove the passive-scalar version of relation \eq{eq01}, and derive some further results in the framework
of the second-order correlation approximation.
We also give an analogous outline of mean-field electrodynamics and present some specific results.
In Sec.~III we
formulate the mean-field concept for the case of non-local relationships between the turbulence-dependent terms
in the mean-field equations and the mean fields,
and we explain the test-field method for the determination of transport coefficients.
In Sec.~IV we then present analytical and numerical results for two simple models
which reflect mean-field properties of irrotational flows.
Finally a discussion of our findings is given in Sec.~V.

\section{Outline of mean-field theories}
\label{Outline}

\subsection{Passive scalar transport}

Let us focus attention on passive scalars $C$ which describe the concentration of, e.g., dust or chemicals
per unit volume of a fluid.
We assume that $C$ satisfies
\EQ
\partial_t C + \nab \cdot (\UU C) - \nab \cdot (\kappa \nab C) = 0 \, ,
\label{eq21}
\EN
where $\UU$ is the fluid velocity and $\kappa$ a diffusion coefficient,
which in general depends on both the mass density $\varrho$ and the temperature $T$.
In the case of an incompressible isothermal fluid, Eq. \eq{eq21} applies with $\kappa$ independent of position
so that $\nab \cdot (\kappa \nab C)$ turns into $\kappa \Delta C$.
We want, however, to include compressible fluids, too.
We may justify \eq{eq21}, e.g., if $C$ describes the concentration of an admixture of light particles
in a compressible isothermal fluid.
The diffusion coefficient is then given by $\kappa = f (T) / \varrho$, with some function $f$;
see \cite{LL}, Chap. \S 11, p.~39.
We expect the validity of \eq{eq21} with some dependency of $\kappa$ on $\varrho$
also in more general cases.
In a flow of such a fluid, its density, even when uniform initially,
will in general become position-dependent in the course of time.
For the sake of simplicity we shall nevertheless ignore any consequence of inhomogeneous density.
Hence, our results are applicable only for either the limited time interval
or the limited velocity amplitude range for which the density inhomogeneity is still negligible.
Overcoming these limitations requires a theory which includes momentum and continuity equations
and is beyond the scope of this paper.

With this in mind we consider, in what follows, $\kappa$ always as independent of position and replace \eq{eq21} by
\EQ
\partial_t C + \nab \cdot (\UU C) - \kappa \Delta C = 0 \, .
\label{eq22}
\EN
We further assume that the fluid motion and therefore also $C$ show turbulent fluctuations,
define mean quantities like $\meanC$ or $\meanUU$ by a proper averaging procedure
which ensures the validity of the Reynolds rules,
and put $C = \meanC + c$ and $\UU = \meanUU + \uu$.
The evolution of $\meanC$ is then governed by
\EQ
\partial_t \meanC + \nab \cdot (\meanUU\, \meanC ) +  \overline{\cal G} - \kappa \Delta \meanC = 0
\label{eq23}
\EN
with
\EQ
\overline{\cal G} = \nab \cdot \meanFFFF \, , \quad \meanFFFF = \overline{\uu c} \, .
\label{eq25}
\EN
For $c$ we have
\EQ
\partial_t c + \nab \cdot \big(\uu \meanC + \meanUU c + (\uu c)'\big) - \kappa \Delta c = 0 \, ,
\label{eq27}
\EN
where $(\uu c)'$ stands for $\uu c - \overline{\uu c}$.
Clearly, $\meanFFFF$ is a functional of $\uu$, $\meanUU$, and $\meanC$ in the sense that
$\meanFFFF$ at a given point in space and time depends in general on
$\uu$, $\meanUU$ and $\meanC$ at other points, too.
This functional is linear in $\meanC$.

Let us, for simplicity, consider the case $\meanUU = {\bm 0}$
and assume that $\uu$ corresponds to homogeneous turbulence.
Until further notice we adopt the assumption that $\meanC$ varies only
weakly in space and time so that $\meanFFFF$, at a given point
in space and time,
can be simply represented as a function of $\meanC$ and its first spatial
derivatives, i.e., $\nab \meanC$, taken just at this point.
We will refer to this assumption as ``perfect scale separation".
We may then conclude that
\EQ
\overline{\cal F}_i = \gamma^{(C)}_i \meanC
    - \kappa_{ij} \frac{\partial \meanC}{\partial x_j} \, ,
\label{eq28}
\EN
with $\gamma^{(C)}_i$ and $\kappa_{ij}$ being coefficients determined by
$\uu$, which are independent of position.\footnote{
Consequently,
a spatially constant $\meanC$ is not influenced by $\uu$ and thus stationary;
nevertheless it causes a fluctuation $c$ that is, for stationary $\uu$, again stationary.
Hence, there is a non-trivial {\em stationary}
solution of \eqref{eq21} with constant average.
For potential flows $\uu=\nabla \phi$ it can be given explicitly as
$C=C_0 \exp(\phi/\kappa)$.
We thank our referee for having made us aware of it.}

Clearly, $\gamma^{(C)}_i$ gives the velocity of advection of $\meanC$,
and $\kappa_{ij}$ is a contribution to the total mean-field diffusivity
tensor, which is then equal to $\kappa \delta_{ij} + \kappa_{ij}$.
From \eq{eq28} we conclude that
\EQ
\overline{\cal G} = \gamma^{(C)}_i \frac{\partial \meanC}{\partial x_i}
    - \kappa_{ij} \frac{\partial^2 \meanC}{\partial x_i \partial x_j} \, .
\label{eq29}
\EN
Of course, $\kappa_{ij}$ may be assumed to be symmetric in $i$ and $j$.
For isotropic turbulence we have $\gamma^{(C)}_i = 0$ and $\kappa_{ij} = \kappa_{\rm t} \delta_{ij}$
with some constant coefficient $\kappat$ so that
\EQ
\meanFFFF = - \kappa_{\rm t} \nab \meanC \, ,
\label{eq30}
\EN
and consequently
\EQ
\overline{\cal G} = - \kappa_{\rm t} \Delta \meanC \, .
\label{eq31}
\EN

Although we have defined $\uu$ as the turbulent velocity, only its statistical symmetry properties
like homogeneity or isotropy have in fact been utilized.
Here and later in this paper, the term `turbulence' should, accordingly, be understood in a wider sense,
including random or even non-random flows with such properties.

In specific calculations often the second-order correlation approximation (SOCA) is used.
It consists in neglecting the term $(\uu c)'$ in
Eq. \eq{eq27} for $c$,
so that
this equation
turns into
\EQ
\partial_t c - \kappa \Delta c = - \nab \cdot (\uu \meanC)  \, .
\label{eq33}
\EN
The applicability of this approximation is restricted to not too large velocities $\uu$.
If we characterize the velocity field again by a typical magnitude $u_{\rm c}$ and by typical length and time scales
$\lambda_{\rm c}$ and $\tau_{\rm c}$, respectively,
we may define the parameter $\qkap = \lambda_{\rm c}^2 / \kappa \tau_{\rm c}$,
which gives the ratio of the free-decay time $\lambda_{\rm c}^2 / \kappa$ to $\tau_{\rm c}$;
further, the P\'eclet number $\Pe = u_{\rm c} \lambda_{\rm c} / \kappa$
and the Strouhal number $\St = u_{\rm c} \tau_{\rm c} / \lambda_{\rm c}$.
Note that $\qkap = \Pe / \St$.
A sufficient condition for the applicability of SOCA in the case $\qkap \gg  1$ reads $\St \ll 1$;
in the case $\qkap \ll 1$ it reads $\Pe \ll 1$.

\subsection{Diffusivity in a special case}

Let us focus attention on homogeneous isotropic turbulence and determine $\kappa_{\rm t}$
in a limiting case.
Since $\kappa_{\rm t}$ depends neither on $\meanC$ nor on position, we may choose simply $\meanC = \GG \cdot \xx$
with a constant $\GG$ so that $\nab \meanC = \GG$, and consider at the end only $\xx = {\bm 0}$.
We represent $\uu$ in the form
\EQ
\uu = \nab \times \ppsi + \nab \phi \, , \quad \nab \cdot \ppsi = 0 \, ,
\label{eq41}
\EN
by a vector potential $\ppsi$ and a scalar potential $\phi$,
and set further
\EQ
\ppsi = \nab \times \cchi \, , \quad \phi = - \nab \cdot \cchi \, ,
\label{eq43}
\EN
where we utilized the freedom in defining the new vector potential $\cchi$ such that
both $\ppsi$ and $\phi$ are now derived from this single quantity.
We then have
\EQ
\uu = - \nab^2 \cchi \, .
\label{eq45}
\EN
We further assume that $\uu$ varies so slowly in time that we may consider it as independent of $t$.
Finally we adopt SOCA so that \eq{eq33} applies.
We may write it in the form
\EQ
\Delta (\kappa c - X) = 0
\label{eq47}
\EN
with
\EQ
X = - \cchi \cdot \GG - 2 \nab \Phi \cdot \GG +  \phi \, \GG \cdot \xx \, ,
\quad \Delta \Phi = \phi \, .
\label{eq49}
\EN
From \eq{eq47} we conclude that
\EQ
c = \frac{1}{\kappa} X + c_0 \, ,
\label{eq51}
\EN
where $c_0$ is some constant.
Calculating then $\meanFFFF$ at $\xx = {\bm 0}$ we obtain
\EQ
\overline{\cal{F}}_i = - \frac{1}{\kappa} \big( \overline{u_i \chi_k}
    + 2 \overline{u_i \partial \Phi / \partial x_k}\big) G_k \, .
\label{eq53}
\EN
Due to the isotropy of the turbulence we have
\EQ
\overline{u_i \chi_k} = \onethird \overline{\uu \cdot \cchi} \,
\delta_{ik} \, , \quad
    \overline{u_i \partial \Phi / \partial x_k} = \onethird \overline{\uu \cdot \nab \Phi} \, \delta_{ik} \, .
\label{eq55}
\EN
Using \eq{eq41} and \eq{eq43} and considering the homogeneity of the turbulence, we find
\EQ
\overline{\uu \cdot \cchi} = \overline{\ppsi^2} + \overline{\phi^2} \, , \quad
\overline{\uu \cdot \nab \Phi} = - \overline{\phi^2} \, .
\label{eq57}
\EN
Consequently we have
\EQ
\meanFFFF = - \frac{1}{3 \kappa} (\overline{\ppsi^2} - \overline{\phi^2}) \GG \, .
\label{eq59}
\EN
Comparing this with \eq{eq30}, we obtain
\EQ
\kappa_{\rm t} = \frac{1}{3 \kappa} (\overline{\ppsi^2} - \overline{\phi^2}) \, ,
\label{eq61}
\EN
a result in full analogy to \eq{eq01}.
For an incompressible flow $\kappat$ can never be negative,
while it can never be positive for an irrotational flow.

\subsection{Relations for transport coefficients}

We consider now homogeneous,
but not necessarily isotropic turbulence
and use a Fourier transformation of the form
\EQ
F (\xx, t) = \int \!\!\! \int \hat{F} (\kk, \omega) \, \exp (\ii \kk \cdot \xx - \ii \omega t) \, \dd^3 k \, \dd \omega \, .
\label{eq81}
\EN
Further we adopt SOCA.
Then standard derivations (see, e.g., \cite{KR80}) yield
\begin{align}
\gamma^{(C)}_i &= - \int \!\!\! \int \frac{\ii k_k}{\kappa k^2 - \ii \omega} \, \hat{Q}_{ik} (\kk, \omega) \, \dd^3 k \, \dd \omega
\label{eq83a}\\
\kappa_{ij} &= \int \!\!\! \int \bigg( \frac{\hat{Q}_{ij} (\kk, \omega) + \hat{Q}_{ji} (\kk, \omega)}{2 (\kappa k^2 - \ii \omega)}
\label{eq83b}\\
&\qquad
     - \frac{2 \kappa \big(\hat{Q}_{ik} (\kk, \omega) k_j + \hat{Q}_{jk} (\kk, \omega) k_i  \big) k_k}{(\kappa k^2 - \ii \omega)^2}
     \bigg) \, \dd^3 k \, \dd \omega \, ,
\nonumber
\end{align}
where $\hat{Q}_{ij} (\kk, \omega)$ is the Fourier transform of the correlation tensor $Q_{ij} (\xxi, \tau)$,
defined by
\EQ
Q_{ij} (\xxi, \tau) =  \overline{u_i (\xx, t) \, u_j (\xx + \xxi, t + \tau)} \, .
\label{eq85}
\EN
Since $Q_{ij} (\xxi, \tau)$ is real, we have $\hat{Q}_{ij} (\kk, \omega) = \hat{Q}_{ij}^\ast (-\kk, -\omega)$,
where the asterisk means complex conjugation.

We recall here Bochner's theorem (see, e.g., \cite{KR80}, Chap. 6),
according to which, for any homogeneous turbulence,
$\hat{Q}_{ij} (\kk, \omega)$ is positive semi-definite, that is,
\EQ
\hat{Q}_{ij} (\kk, \omega) \, X_i \, X^*_j \geq 0
\label{eq91}
\EN
for any complex vector $\XX$.

Assume first incompressible turbulence, that is, \mbox{$\nab \cdot \uu = 0$}.
Then we have
\EQ
\hat{Q}_{ij} k_j = 0 \, , \quad \hat{Q}_{ij} k_i = 0 \, .
\label{eq101}
\EN
In this case, \eq{eq83a} yields
$\gamma^{(C)}_i = 0$ (even if the flow is not isotropic),
and \eq{eq83b} turns into
\EQ
\kappa_{ij} = \frac{1}{2} \int \!\!\! \int \frac{1}{\kappa k^2 - \ii \omega}
        \left( \hat{Q}_{ij} (\kk, \omega) + \hat{Q}_{ji} (\kk, \omega) \right) \, \dd^3 k \, \dd \omega \, .
\label{eq103}
\EN
From \eq{eq91} and \eq{eq103} we may conclude that $\kappa_{ij}$ is positive semi-definite.
If the flow is statistically isotropic we have $\kappa_{ij} = \kappa_{\rm t}  \delta_{ij}$
and we may conclude that $\kappa_{\rm t}$ is non-negative.

Assume next irrotational turbulence, that is,
$\uu = \nab \phi$ with any potential $\phi$.
Then we have
\EQ
\hat{Q}_{ij} (\kk, \omega) = k_i k_j \hat{R} (\kk, \omega)
\label{eq111}
\EN
with some real function $\hat{R}$
related to $\phi$.
Owing to \eq{eq91}, $\hat{R}$ must be non-negative
(cf. \cite{KR80}, Chap. 6).
With \eq{eq83a}, \eq{eq83b} and \eq{eq111} we find
\EQA
\gamma^{(C)}_i &=& - \int \!\!\! \int \frac{\ii k_i \hat{R} (\kk, \omega) k^2}{\kappa k^2 - \ii \omega} \, \dd^3 k \, \dd \omega
\label{eq113a}\\
\kappa_{ij} &=& - \int \!\!\! \int \frac{(3 \kappa k^2 + \ii \omega) k_i k_j \hat{R} (\kk, \omega)}{(\kappa k^2 - \ii \omega)^2}
    \, \dd^3 k \, \dd \omega \, .
\label{eq113b}
\ENA
For statistically isotropic flows $\hat{R}$ depends only via $k$ on $\kk$.
Hence, we obtain as expected
$\gamma^{(C)}_i = 0$ and
\EQ
\kappa_{\rm t} = - \frac{1}{3} \int \!\!\! \int \frac{(3 \kappa k^2 + \ii \omega) k^2 \hat{R} (k, \omega)}{(\kappa k^2 - \ii \omega)^2}
    \, \dd^3k \, \dd \omega \, .
\label{eq115}
\EN
Assume now in addition that the variations of $\uu$ in time are slow.
Then $\hat{R}$ is markedly different from zero only for very small $\omega$.
Consequently, $\kappa_{\rm t}$ is non-positive.

\subsection{Magnetic-field transport}
\label{magnetic}

We now consider a magnetic field $\BB$ in a homogeneous electrically conducting fluid
and assume that it is governed by
\EQ
\partial_t \BB - \nab \times (\UU \times \BB) - \eta \nab^2 \BB = {\bm 0} \, , \quad \nab \cdot \BB = 0 \, ,
\label{eq161}
\EN
with $\UU$ being again the velocity and $\eta$ the magnetic diffusivity of the fluid.
Focusing attention on
a turbulent situation, we define again mean fields, in particular $\meanBB$ and $\meanUU$,
and put $\BB = \meanBB + \bb$ and $\UU = \meanUU + \uu$.
Then we have
\EQ
\partial_t \meanBB - \nab \times (\meanUU \times \meanBB + \meanEMF) - \eta \nab^2 \meanBB = {\bm 0} \, , \quad \nab \cdot \meanBB = 0 \, , \label{eq163}
\EN
where
\EQ
\meanEMF = \overline{\uu \times \bb}
\label{eq165}
\EN
and
\EQA
\partial_t \bb - \nab \times \big(\meanUU \times \bb + \uu \times \meanBB + (\uu \times \bb)'\big) \!\!&-&\!\! \eta \nab^2 \bb = {\bm 0} \, ,
\nonumber\\
&&\!\!\!\!\!\!\!\!\!\! \nab \cdot \bb = 0 \, .
\label{eq167}
\ENA
Here $(\uu \times \bb)'$ means $\uu \times \bb - \overline{\uu \times \bb}$.
The mean electromotive force $\meanEEEE$ due to the fluctuations $\uu$ and $\bb$
is a functional of $\uu$, $\meanUU$, and $\meanBB$, which is linear in $\meanBB$.

Let us restrict ourselves again to $\meanUU = {\bm 0}$.
Assuming perfect scale separation, defined analogously to the passive scalar case considered before,
we may conclude that
\EQ
\meanemf_i = a_{ij} \meanB_j - \eta_{ij} (\nab \times \meanBB)_j - c_{ijk} (\nab \meanBB)^{\rm s}_{jk} \, ,
\label{eq169}
\EN
where $(\nab \meanBB)^{\rm s}_{jk} = \half (\partial \meanB_j / \partial x_k + \partial \meanB_k / \partial x_j)$.
Here $a_{ij}$, $\eta_{ij}$ and $c_{ijk}$ are quantities determined by $\uu$.
(Instead of the traditional $b_{ijk}$ we use here $\eta_{ij} = \half \, b_{imn} \epsilon_{jmn}$
and $c_{ijk} = - \half (b_{ijk} + b_{ikj})$.)\footnote{
In analogy to what was noted for the passive scalar case,
stationary solutions of \eq{eq161} with constant mean parts of $\BB$ are conceivable.
}

In this context SOCA consists in dropping the term $(\uu \times \bb)'$ in \eq{eq167} so that
\EQ
\partial_t \bb - \eta \nab^2 \bb = \nab \times (\uu \times \meanBB) \, , \quad \nab \cdot \bb = 0 \, .
\label{eq171}
\EN
Sufficient conditions under which this applies are again analogous to those explained below \eq{eq33}.
We have only to replace the parameter $q_\kappa$ by $q_\eta = \lambda_c^2 / \eta \tau_{\rm c}$
and the P\'eclet number $\Pe$ by the magnetic Reynolds number $\Rm = u_{\rm c} \lambda_{\rm c} / \eta$.
Note that $\qeta = \Rm / \St$.

The relevant relations for $a_{ij}$, $\eta_{ij}$, and $c_{ijk}$, derived under SOCA for homogeneous turbulence,
are given in the Appendix~\ref{appA}.
Let us restrict ourselves here to homogeneous non-helical turbulence.
Then the correlation tensor $\hat{Q}_{ij}$ may not contain any pseudo-scalar or any other pseudo-quantity.
As a consequence, the symmetric part of $a_{ij}$ and the antisymmetric part of $\eta_{ij}$ are equal to zero,
and we have
\EQA
a_{ij} &\!=\!& \epsilon_{ijk} \gamma^{(B)}_k \, ,
\nonumber\\
\gamma^{(B)}_i &\!=\!& \frac{1}{2} \int \!\!\! \int \ii k_k \frac{\hat{Q}_{ik} (\kk, \omega) + \hat{Q}_{ki} (\kk, \omega)
    }{\eta k^2 - \ii \omega} \, \dd^3k \, \dd \omega \, ,
\label{eq173}
\ENA
and
\begin{align}
\hspace*{-5mm}\eta_{ij}& = \frac{1}{2} \int \!\!\! \int \!\! \left( \frac{[2 \delta_{ij} \delta_{kl} - (\delta_{ik} \delta_{jl}
    + \delta_{jk} \delta_{il})]}{2(\eta k^2 - \ii \omega)}\right.
\nonumber\\
&\hspace*{-1.5mm}-\left.\frac{\eta\, [2 \delta_{ij} k_k - (k_i \delta_{jk} + k_j \delta_{ik})]\, k_l
     }{(\eta k^2 - \ii \omega)^2} \!\right)\!\hat{Q}_{kl} (\kk, \omega) \, \dd^3 k \, \dd \omega \, .
\label{eq175}
\end{align}
Moreover, $c_{ijk}$ is equal to zero.

Consider now incompressible turbulence, for which \eq{eq101} applies.
Then we have, even in the anisotropic case, $\gamma^{(B)}_i = 0$
[see also \cite{KR80}, Chap. 7.1, statement (i)].
Furthermore,
\EQ
\eta_{ij} = \frac{1}{4} \int \!\!\! \int \frac{[2 \delta_{ij} \delta_{kl} - (\delta_{ik} \delta_{jl} + \delta_{jk} \delta_{il})]
     \hat{Q}_{kl} (\kk, \omega)}{\eta k^2 - \ii \omega} \, \dd^3 k \, \dd \omega \, ,
\label{eq179}
\EN
which, together with \eq{eq91}, implies that $\eta_{ij}$ is positive semi-definite.
If the turbulence is in addition isotropic, we have
$\eta_{ij} = \eta_{\rm t} \delta_{ij}$ with
\EQ
\eta_{\rm t} = \frac{1}{3} \int \!\!\! \int \frac{\hat{Q}_{kk} (\kk, \omega)}{\eta k^2 - \ii \omega} \, \dd^3 k \, \dd \omega \, .
\label{eq181}
\EN
Like $\kappa_{\rm t}$, $\eta_{\rm t}$ also has to be non-negative
[see also \cite{KR80}, Chap. 7.4, Eq. (7.47)].

Consider next irrotational turbulence, for which \eq{eq111} applies.
Then,
\begin{align}
\hspace*{-2.5mm}\gamma^{(B)}_i &= \int \!\!\! \int \frac{\ii k_i k^2 \hat{R} (\kk, \omega)}{\eta k^2 -\ii \omega} \, \dd^3k \, \dd \omega \, ,
\label{eq183a}\\
\hspace*{-2.5mm}\eta_{ij} &= - \frac{1}{2} \int \!\!\! \int\! \frac{( \eta k^2 + \ii \omega) (k^2 \delta_{ij} - k_i k_j) \hat{R} (\kk, \omega)
    }{(\eta k^2 - \ii \omega)^2}
\label{eq183b}    \,\dd^3 k \, \dd \omega \, .
\end{align}
If the variations of $\uu$ in time are slow,
$\hat{R}$ is markedly different from zero only for small $\omega$.
Then it can be readily shown that $\eta_{ij}$ is negative semi-definite.
In the isotropic case, $\hat{R}$ depends, as already noted above, only via $k$ on $\kk$.
Therefore we have,
independent of the time behavior of $\uu$,
$\gamma^{(B)}_i = 0$ and
\EQ
\eta_{\rm t} = - \frac{1}{3} \int \!\!\! \int \frac{(\eta k^2 + \ii \omega) \hat{R} (k, \omega) k^2
    }{(\eta k^2 - \ii \omega)^2} \, \dd^3 k \, \dd \omega \, .
\label{eq185}
\EN
If then the time-variations of $\uu$ are slow,
$\eta_{\rm t}$ has to be non-positive
[see also \cite{KR80}, Chap. 7, Eq. (7.51)].

\section{Generalizations and test-field procedure}
\label{Generalizations}

\subsection{Lack of scale separation}

In applications, the assumption of perfect scale separation, used so far,
might be violated; see, e.g., \cite{Chatterjee}.
We now relax it.
Considering first again the passive scalar case we admit now
a dependence of $\meanFFFF$, at a given point in space,
on $\meanC$ (or its derivatives) at other points,
that is, we admit a non-local connection between $\meanFFFF$ and $\meanC$.
For the sake of simplicity, however, we assume until further notice that $\meanFFFF$, at a given time,
is only connected with $\meanC$ (or its derivatives) at the same time,
that is, we remain with an instantaneous connection between $\meanFFFF$ and $\meanC$.
Again, we restrict ourselves to homogeneous turbulence.

We further assume here, again for simplicity,
that mean fields are defined as averages over all $x$ and $y$.
Hence, they depend on $z$ and $t$ only.

In what follows it is then sufficient to consider the $z$ component of $\meanFFFF$ only.
As a straightforward generalization of the relation for $\meanFFF_z$
contained in \eq{eq28}, with derivatives with respect to $z$ only, we now write
\begin{align}
\meanFFF_z (z,t) &= \int \bigg(\gamma_z^{(C)}(\zeta) \, \meanC (z - \zeta, t)
\nonumber\\
& \qquad \quad
    - \kappa_{zz} (\zeta) \frac{\partial \meanC (z - \zeta, t)}{\partial z} \bigg) \, \dd \zeta \, ,
\label{eq201}
\end{align}
with two functions $\gamma_z^{(C)} (\zeta)$ and $\kappa_{zz} (\zeta)$, which
are assumed to be symmetric in $\zeta$,
and with the integration being
over all $\zeta$.
(In the case of inhomogeneous turbulence, $\gamma_z^{(C)}$ and $\kappa_{zz}$
would also depend on $z$.)
With the specifications $\gamma_z^{(C)} (\zeta) = \gamma_z^{(C)} \delta(\zeta)$ and $\kappa_{zz} (\zeta) = \kappa_{zz} \delta(\zeta)$,
where $\gamma_z^{(C)}$ and $\kappa_{zz}$ on the right hand sides are understood as constants,
we return just to the relation for $\meanFFF_z$ given by \eq{eq28}.
Utilizing integrations by parts, we may rewrite \eq{eq201} as
\EQ
\meanFFF_z (z, t) = \int \Gamma (\zeta) \meanC (z - \zeta, t) \, \dd \zeta
\label{eq203}
\EN
with
\EQ
\Gamma (\zeta) = \gamma_z^{(C)} (\zeta) - \frac{\partial \kappa_{zz} (\zeta)}{\partial \zeta} \, .
\label{eq205}
\EN

In what follows, it is convenient to work with a Fourier transformation
defined by
\EQ
F (\zeta) = \frac{1}{2 \pi} \int \tilde{F} (k) \exp(\ii k \zeta) \, \dd k \, .
\label{eq207}
\EN
[Apart from the fact that here only a function of the single variable $\zeta$ is considered,
this definition differs from \eq{eq81} also by the factor $1/ 2 \pi$
on the right-hand side.]
\EEq{eq203} is then equivalent to
\EQ
\meanFFF_z (z, t) = \frac{1}{2 \pi} \int \tilde{\Gamma} (k) \tilde{\meanC} (k, t) \exp(\ii k z) \, \dd k \, ,
\label{eq209}
\EN
and \eq{eq205} implies
\EQA
\tilde{\gamma}_z^{(C)} (k) &=& \Re (\tilde{\Gamma} (k))
\nonumber\\
\tilde{\kappa}_{zz} (k) &=& - k^{-1} \Im (\tilde{\Gamma} (k)) \, .
 \label{eq211}
\ENA
For $\gamma_z^{(C)}$ and $\kappa_{zz}$ on the right-hand sides of \eq{eq28} and \eq{eq29}
we have then
\EQ
\gamma_z^{(C)} = \tilde{\gamma}_z^{(C)} (0) \, , \quad \kappa_{zz} = \tilde{\kappa}_{zz} (0) \, .
\label{eq213}
\EN

Let us now admit that $\meanFFF_z$, at a given time, depends on $\meanC$
(and its spatial derivatives)
not only at this but also at earlier times.
This non-instantaneous connection between $\meanFFF_z$ and $\meanC$
can be described as a {\em memory effect};
see, e.g., \cite{HB09}.
We then have to generalize \eq{eq201} such that
\EQA
\meanFFF_z (z) &=& \int \!\!\! \int \bigg(\gamma_z^{(C)} (\zeta, \tau) \, \meanC (z - \zeta, t - \tau)
\nonumber\\
&& - \kappa_{zz} (\zeta, \tau) \frac{\partial \meanC (z - \zeta, t - \tau)}{\partial z} \bigg) \, \dd \zeta \, \dd \tau
\label{eq231}
\ENA
with $\gamma_z^{(C)}$ and $\kappa_{zz}$ symmetric in $\zeta$
and equal to zero for $\tau < 0$;
the integration is then over all $\zeta$ and $\tau \geq 0$.
It is straightforward to generalize the relations \eq{eq203} to \eq{eq213} in that sense.
Then, Fourier transforms with respect to $\zeta$ and $\tau$ occur, and $\tilde{\gamma}^{(C)}_z$ and $\tilde{\kappa}_{zz}$
depend not only on $k$ but also on an additional variable $\omega$.

The generalizations explained here can easily be extended to the magnetic case discussed in Sec.~\ref{magnetic}.
Then, $\gamma^{(B)}_z$, $\eta_{xx}$, and $\eta_{yy}$ occur as functions of $\zeta$, or of $\zeta$ and $\tau$, and
their Fourier transforms $\tilde{\gamma}^{(B)}_z$, $\tilde{\eta}_{xx}$, and $\tilde{\eta}_{yy}$ as functions
of $k$ , or of $k$ and $\omega$.

\subsection{Test-field procedure}
\label{subsIIIB}

In Sec.~\ref{Outline}
we have presented results for quantities like $\gamma^{(C)}_i$ or $\kappa_{ij}$ which apply only under SOCA.
As soon as we are able to solve
equations like \eq{eq27}, e.g., numerically, we may determine these quantities,
or $\tilde{\gamma}_z^{(C)}$ and $\tilde{\kappa}_{zz}$ introduced in the preceding section, also beyond this approximation.
A proper tool for that is the test-field method, first developed in mean-field electrodynamics
\cite{SRSRC07,BRS08}.
We apply the ideas of this method here first to the passive scalar case.
As in the preceding section we assume again that the mean fields are defined by averaging over all $x$ and $y$ and
relax
spatial scale separation, but ignore at first scale separation in time, that is, the memory effect.

Suppose that we have solved \eq{eq27} for two different {\it test-fields} $\meanC$, say
\EQ
\meanC^{\,\rm c} = C_0 \cos kz \quad \mbox{and} \quad \meanC^{\,\rm s} = C_0 \sin kz
\label{eq221}
\EN
with given $C_0$ and $k$, and calculated the corresponding $\meanFFF_z$,
say $\meanFFF_z^{\,\rm c}\!(z)$ and $\meanFFF_z^{\,\rm s}\!(z)$.
Specifying \eq{eq201} to $\meanFFF_z^{\,\rm c}$ and $\meanFFF_z^{\,\rm s}$ and considering that,
due to \eq{eq207} and the assumed symmetry of $\gamma_z^{(C)} (\zeta)$ and $\kappa_{zz} (\zeta)$ in $\zeta$,
\EQA
\int \gamma_z^{(C)} (\zeta) \, \cos k \zeta \, \dd \zeta &=& \tilde{\gamma}_z^{(C)} (k) \, ,
\nonumber\\
\int \kappa_{zz} (\zeta) \, \cos k \zeta \, \dd \zeta &=& \tilde{\kappa}_{zz} (k) \, ,
\label{eq223}
\ENA
we find
\begin{alignat}{2}
\meanFFF_z^{\,\rm c}\! (z) &=  C_0 \big( &&\tilde{\gamma}_z^{(C)} (k) \, \cos k z + \tilde{\kappa}_{zz}(k) \, k \, \sin k z \big)
\nonumber\\
\meanFFF_z^{\,\rm s}\! (z) &=  C_0 \big( &&\tilde{\gamma}_z^{(C)} (k) \, \sin k z - \tilde{\kappa}_{zz}(k)\, k \, \cos k z \big) \, .
\label{eq225}
\end{alignat}
This in turn leads to
\begin{alignat}{2}
\tilde{\gamma}_z^{(C)} (k) &= \frac{1}{C_0} &&\big(\meanFFF_z^{\,\rm c}\! (z) \, \cos k z
    + \meanFFF_z^{\,\rm s}\! (z) \, \sin k z \big)
\nonumber\\
\tilde{\kappa}_{zz} (k) &= \frac{1}{C_0 k}
 &&\big(\meanFFF_z^{\,\rm c}\! (z) \, \sin k z
    - \meanFFF_z^{\,\rm s}\! (z) \, \cos k z \big) \, .
\label{eq227}
\end{alignat}
Note that, although
constituents of the right-hand sides depend on $z$, the left-hand sides do not.

If the memory effect is taken into account, Eq. \eq{eq27} has to be solved
with time-dependent test-fields $\meanC$.
Let us define such fields by multiplying the right-hand sides in \eq{eq221} by a factor $e^{\ii \omega t}$.
Integrate then the relevant equations numerically with any initial condition
until all transient parts of the solutions have disappeared.
For steady flows, the remaining solutions then show the same harmonic time variation
as the test fields (approximately possible also for unsteady flows).
That is, the same time-dependent factors occur on both sides
of the equations analogous to \eq{eq225} and can be removed.
These equations then allow the determination of
$\tilde{\gamma}^{(C)}_z (k, \omega)$ and $\tilde{\kappa}_{zz} (k, \omega)$,
that is, the Fourier transforms of $\gamma^{(C)}_z (\zeta, \tau)$ and $\kappa_{zz} (\zeta, \tau)$.
Of course, $\tilde{\gamma}^{(C)}_z (k, \omega)$ and $\tilde{\kappa}_{zz} (k, \omega)$ are in general complex.
We may also replace the factor $e^{\ii \omega t}$ by $e^{\sigma t}$ with a complex $\sigma$.
Instead of the Fourier transformation with respect to time
we have then to use a Laplace transformation.

A test-field procedure, as described here for passive scalars, can
also be established for the magnetic case as discussed
in Sec.~\ref{magnetic}.
It allows then the calculation of quantities like $\gamma^{(B)}_z$,  $\eta_{xx}$, and $\eta_{yy}$
or their Fourier or Laplace transforms.
Such procedures have already been used elsewhere (e.g., \cite{HB09,BRS08}).

For the numerical computations presented below we use the
{\sc Pencil Code} \cite{PC}, where the test-field
methods both for passive scalars and for magnetic fields
have already been implemented \cite{BSV09}.
All results presented in this paper have been obtained with
a version of the code compatible with revision {\tt 16408}.

\section{Examples and Illustrations}

\subsection{Three-dimensional flow}

In an attempt to model properties of homogeneous isotropic irrotational
turbulence we wish to consider first a steady potential flow.
Thus, we choose
\begin{align}
\hspace*{-1mm}\uu &= \nab \phi
\label{eq251}\\
\hspace*{-1mm}\phi &= \frac{u_0}{k_0} \cos k_0(x + \chi_x) \, \cos k_0(y + \chi_y) \, \cos k_0(z + \chi_z). \hspace{-1mm}
\label{eq253}
\end{align}
Here, $u_0$ and $k_0$ are positive constants
and $\chi_x$, $\chi_y$, and $\chi_z$ are understood as random phases.
Of course, this steady flow must lead to growing inhomogeneities of the mass density.
Therefore the applicability of our results is restricted to a limited time range,
see the discussion below \eq{eq21}.

Starting from original fields $C$ and $\BB$, which may depend on $x$, $y$, $z$, and $t$, and also on $\chi_x$, $\chi_y$, and $\chi_z$,
we define mean fields $\meanC$ and $\meanBB$
by averaging over all $x$ and $y$ and, in addition,
over $\chi_z$.
Consequently, mean fields no longer depend on $x$, $y$, or $\chi_z$,
but they may depend on $z$ and $t$.
Clearly, the Reynolds averaging rules apply exactly.
For mean quantities determined by $\uu$ only,
averaging over $x$ and $y$ is equivalent to averaging over $\chi_x$ and $\chi_y$.
Therefore, such quantities can also be understood
as averages over $\chi_x$, $\chi_y$, and $\chi_z$.
Clearly, $\overline{\uu^2}$ is independent of $x$, $y$,
and also of $z$.

From \eq{eq251} and \eq{eq253}, we conclude that
\EQ
u_{\rm rms} = \frac{1}{2} \sqrt{\frac{3}{2}} \,u_0
\label{eq255}
\EN
and we define a wave number $\kf$
of $\uu$ by
\EQ
\kf = \sqrt{3} k_0 \, .
\label{eq257}
\EN
In what follows, $\kappa_{\rm t}$ and $\eta_{\rm t}$, as well as
$\tilde{\kappa}_{\rm t}$ and $\tilde{\eta}_{\rm t}$,
will be expressed in units of $\kappa_{\rm t0}$ and $\eta_{\rm t0}$,
given by
\EQ
\kappa_{\rm t0} = \eta_{\rm t0} = \frac{u_{\rm rms}}{3 \kf} \, .
\label{eq259}
\EN
Furthermore, we define the P\'eclet number $\Pe$ and the magnetic Reynolds
number $\Rm$ by
\EQ
\Pe = \frac{u_{\rm rms}}{\kappa \kf} \, , \quad \Rm = \frac{u_{\rm rms}}{\eta \kf} \, .
\label{eq261}
\EN

Calculations in the framework of SOCA under the assumption of perfect scale separation yield
\EQA
\kappa_{\rm t} &=& - \kappa_{\rm{t}0} \, \Pe \, ,
\nonumber\\
\eta_{\rm t} &=& - \eta_{\rm{t}0} \, \Rm \, .
\label{eq265}
\ENA
Clearly, $\kappa_{\rm t}$ and $\eta_{\rm t}$ are non-positive.
If scale separation is, in the sense of \eq{eq201}, relaxed, we obtain
\EQA
\tilde{\kappa}_{\rm t} (k) &=& - \kappa_{{\rm t}0} \, \Pe \;\; f (k/\kf) \, ,
\nonumber\\
\tilde{\eta}_{\rm t} (k) &=&  - \eta_{{\rm t}0} \, \Rm \, f (k/\kf) \, ,
\label{eq267}\\
f (v) &=& \frac{1 - v^2}{1 + (2 / 3) v^2 + v^4} \, .
\nonumber
\ENA
In all following discussions we consider $k$ as positive.
Like $\kappat$ and $\etat$, also $\tilde{\kappa}_{\rm t}$ and $\tilde{\eta}_{\rm t}$ are negative
as long as $k/\kf < 1$.

In what follows, we present results for the quantities $\tilde{\kappa}_{\rm t}$ and $\tilde{\eta}_{\rm t}$
obtained by the test-field procedure described in Sec. \ref{subsIIIB}, utilizing numerical integrations
of Eq. \eq{eq27} for $c$ or Eq. \eq{eq167} for $\bb$.
Averaging over $\chi_z$ was performed by averaging over $z$.

\begin{figure}[t!]\begin{center}
\includegraphics[width=\columnwidth]{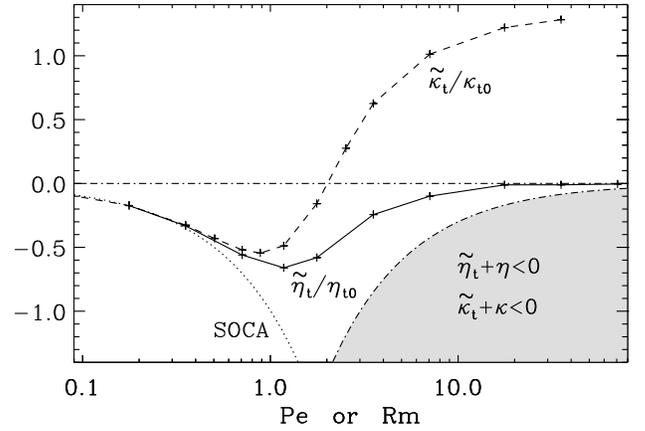}
\end{center}\caption[]{
$\tilde{\kappa}_{\rm t}/\kappa_{\rm t0}$ and
$\tilde{\eta}_{\rm t}/\eta_{\rm t0}$
as functions of $\Pe$ and $\Rm$, respectively,
for the model given by \eqref{eq251} and \eqref{eq253};
$k/\kf = 1 / 10 \sqrt{3} \approx 0.06$.
The dotted line on the lower left gives the SOCA result and the
shaded area on the lower right marks the regime where the total
diffusivities would become negative.
}\label{pkap}\end{figure}

Figure~\ref{pkap} shows $\tilde{\kappa}_{\rm t} / \kappa_{\rm t0}$ as well as $\tilde{\eta}_{\rm t} / \eta_{\rm t0}$
for a small value of $k / \kf$, at which these quantities should be very close to $\kappa_{\rm t} / \kappa_{\rm t0}$
and $\eta_{\rm t} / \eta_{\rm t0}$, as functions of $\Pe$ and $\Rm$, respectively.
(These values could also be obtained with a test field that is independent of $z$ and another one linear in $z$.)
In agreement with the results presented in
Sec.~\ref{Outline} and with \eq{eq265},
$\kappa_{\rm t}$ and $\eta_{\rm t}$ are negative for not too large values of $\Pe$ and $\Rm$, respectively.
Remarkably the functions $\tilde{\kappa}_{\rm t} (\Pe)$ and $\tilde{\eta}_{\rm t} (\Rm)$
coincide formally for small values of $\Pe$ and $\Rm$ only,
but are otherwise clearly different from each other.
In particular, $\tilde{\eta}_{\rm t}$ remains negative,
at least for $\Rm\le70$,
while $\tilde{\kappa}_{\rm t}$ becomes positive for $\Pe\ga2$.
The total diffusivities, $\eta+\tilde{\eta}_{\rm t}$ and
$\kappa+\tilde{\kappa}_{\rm t}$, are always found to be positive.

\begin{figure}[t!]\begin{center}
\includegraphics[width=\columnwidth]{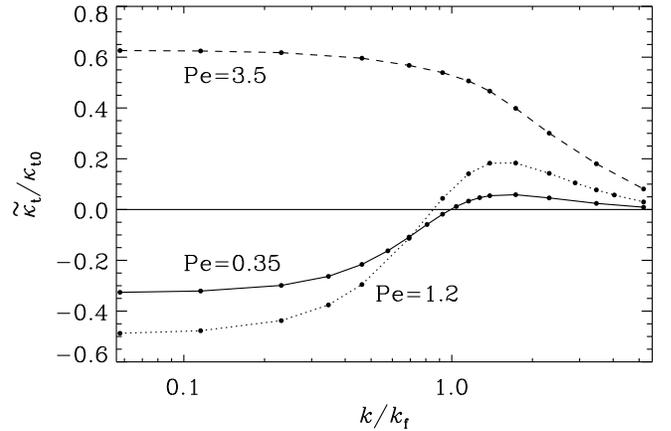}
\end{center}\caption[]{
$\tilde{\kappa}_{\rm t}/\kappa_{\rm t0}$ versus $k/\kf$
for some values of $\Pe$.
}\label{pk5}\end{figure}

\begin{figure}[t!]\begin{center}
\includegraphics[width=\columnwidth]{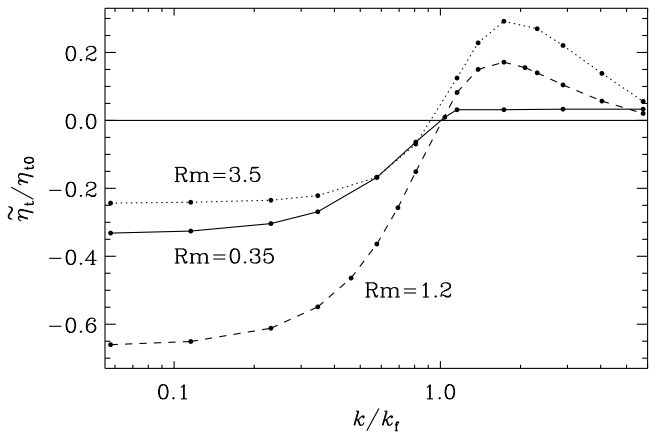}
\end{center}\caption[]{
$\tilde{\eta}_{\rm t}/\eta_{\rm t0}$ versus $k/\kf$ for some values of $\Rm$.
}\label{pk_dep}\end{figure}

Figures~\ref{pk5} and \ref{pk_dep} show examples of the dependence
of $\tilde{\kappa}_{\rm t} / \kappa_{\rm t0}$ and $\tilde{\eta}_{\rm t} / \eta_{\rm t0}$ on $k / \kf$.
Again, $\tilde{\kappa}_{\rm t}$ and $\tilde{\eta}_{\rm t}$ with $\Pe = 0.35$ and $\Rm = 0.35$, respectively,
that is, in the validity range of SOCA, take coinciding negative values in the limit of small $k / \kf$.
However, $\tilde{\kappa}_{\rm t}$ and $\tilde{\eta}_{\rm t}$ become positive for large values of $k / \kf$,
regardless of the values of $\Pe$ and $\Rm$.
The dependence of $\tilde{\kappa}_{\rm t}$ on $\Pe$ and that of $\tilde{\eta}_{\rm t}$ on $\Rm$  are in general
clearly different from each other.

The results regarding negative contributions of $\kappa_{\rm t}$ to the
mean-field diffusivity for passive scalars,
or negative contributions of $\eta_{\rm t}$ to the magnetic mean-field diffusivity,
have been found under the assumption that the velocity $\uu$ is steady or varies only weakly in time.
In order to see the influence of the variability of $\uu$ we consider now a {\em renovating flow}.
It is assumed that, during some time interval, a steady flow as given by \eq{eq253} with some values of $\chi_x$, $\chi_y$
and $\chi_z$ exists, and likewise in the following interval,
but with randomly changed $\chi_x$, $\chi_y$ and $\chi_z$, and so forth.
Hence, there is no correlation between the flows in the different intervals.
It is further assumed that all intervals are equally long.
Denoting their durations by $\tau$, we define now the dimensionless parameters
\EQ
\qkap = (\kappa \kf^2 \tau)^{-1} \, , \quad \qeta = (\eta \kf^2 \tau)^{-1} \, .
\label{eq269}
\EN
Steadiness of the velocity corresponds to $\qkap = \qeta = 0$.

\begin{figure}[t!]\begin{center}
\includegraphics[width=\columnwidth]{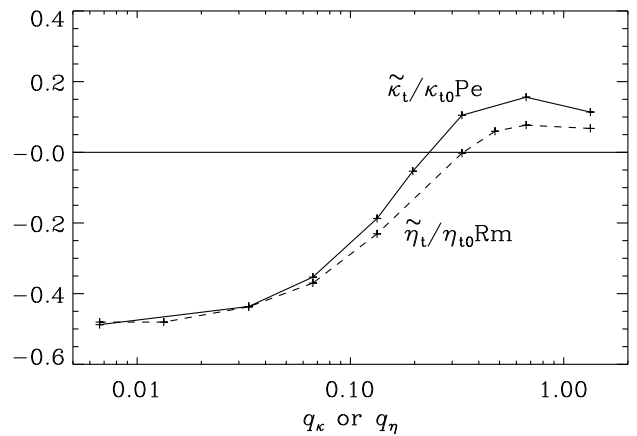}
\end{center}\caption[]{
$\tilde{\kappa}_{\rm t}/\kappatz \Pe$ and $\tilde{\eta}_{\rm t}/\etatz \Rm$ versus
$q_\kappa$ or $q_\eta$, respectively,
for a renovating flow with $\urms\kf\tau = 5.3$ and $k/\kf=1/\sqrt{3}\approx0.6$.
In the calculations,
$\tau$ was held constant.
Consequently, $q_\kappa=\Pe/5.3$ and $q_\eta=\Rm/5.3$.
}\label{pboth}\end{figure}

Figure~\ref{pboth} shows the dependency of $\tilde{\kappa}_{\rm t}/\kappatz \Pe$
on $\qkap$ and that of $\tilde{\eta}_{\rm t}/\etatz \Rm$ on $\qeta$
for $k/\kf=1/\sqrt{3}\approx0.6$.
We see that $\tilde{\kappa}_{\rm t}$ and $\tilde{\eta}_{\rm t}$ are no longer negative
if $\qkap$ and $\qeta$ exceed $0.2$ and $0.3$, respectively.

\subsection{Plane wave-like flow}

With the idea of establishing a simple model reflecting features
of homogeneous {\em anisotropic}
turbulence, we remain with \eq{eq251}, that is $\uu = \nab \phi$, but replace \eq{eq253} by
\EQ
\phi = \frac{u_0}{k_0} \cos \big(k_0 (s x + z) - \omega_0 t - \chi\big)\,.
\label{eq272}
\EN
If $s = 0$, the velocity $\uu$ corresponds to a sound wave traveling in
the $z$ direction,
with wavelength and frequency determined by $k_0$ and $\omega_0$
and with a phase angle $\chi$.
We assume, for simplicity, $k_0 > 0$ and $\omega_0 \geq 0$
so that the wave travels in the positive $z$ direction.
If we admit non-zero values of $s$, the wave propagates no longer in the $z$ direction,
but in the direction of the vector $(s, 0, 1)$.
For $\omega_0 = 0$ the velocity $\uu$ does not depend on time, that is, we have a ``frozen-in" wave.

Similarly to the preceding example we define mean fields here by averaging over all $x$ and $y$ and,
if the original field depends on $\chi$, also over $\chi$.
Then, mean fields may depend only on $z$ and $t$.
Again, the Reynolds averaging rules apply exactly.
If an original field is determined by $\uu$ only and $s$ is unequal to zero,
averaging over $x$ is equivalent to averaging over $\chi$.

Instead of \eq{eq255} we now have
\EQ
u_{\rm rms} = u_0\sqrt{\frac{s^2+1}{2}} \, ,
\label{eq273}
\EN
and instead of \eq{eq257} and \eq{eq259} we put
\EQ
\kf = k_0
,\quad\;
\kappa_{\rm t0} = \eta_{\rm t0} = \frac{u_{\rm rms}}{\kf} \, ,
\label{eq277}
\EN
and we define $\Pe$ and $\Rm$ again according to \eq{eq261}.
Finally we set
\EQ
\qkap = \frac{\omega_0}{\kappa k_0^2} \, , \quad
\qeta = \frac{\omega_0}{\eta k_0^2} \, .
\label{eq279}
\EN

Due to the definition of mean fields, which implies that $\meanC$ and $\meanBB$ do not depend on $x$ and $y$,
we also have $(\nab \times \meanBB)_z = 0$.
In addition, we assume that $\meanB_z = 0$.

Adopting SOCA and assuming again perfect scale separation, we find
\EQA
\gamma^{(C)}_z &=& u_{\rm rms} \, \Pe  \;\; g (s, \qkap)
    \nonumber \, ,\\
\gamma^{(B)}_z &=& u_{\rm rms} \, \Rm \, g (s, \qeta)
\label{eq285} \, ,\\
g (s, q) &=& \frac{(1+s^2) q}{(1+s^2)^2 + q^2}
\nonumber  \, ,
\ENA
and
\begin{alignat}{2}
\kappa_{zz} &= \,&-& \kappa_{{\rm t}0} \, \Pe  \;\; h (s, \qkap)
\nonumber \, ,\\
\eta_{xx} &= \eta_{yy} = \,&-& \eta_{{\rm t}0} \, \Rm \, h (s, \qeta) \, ,
\label{eq287}
\end{alignat}
\EQ
h (s, q) = \frac{(1 + s^2) \big((1 + s^2)^2 - 3 q^2\big)}{\big((1 + s^2)^2 + q^2\big)^2}  \, .
\nonumber
\EN
In the case $\qkap = \qeta = 0$, that is, for frozen-in
waves, the $\gamma^{(C)}_z$ and $\gamma^{(B)}_z$  vanish.
This is due to the fact that, then, there is no preference for the positive or negative $z$ direction.
For $\qkap \not= 0$, however, $\gamma^{(C)}_z$ is positive so that $\meanC$ is advected in the positive $z$ direction.
Further, $\kappa_{zz}$ is negative for not too large $\qkap$, but it becomes positive for larger $\qkap$.
This applies analogously with $\qeta$, $\gamma^{(B)}_z$, and $\eta_{xx} = \eta_{yy}$.

Relaxing perfect spatial scale separation and assuming that $\meanC$ and $\meanBB$ vary in time
as $\exp (\sigma t)$ with a real $\sigma$, we find further
\begin{align}
\tilde{\gamma}^{(C)}_z &= u_{\rm rms} \, \Pe \;\;\, g (s, \qkap, k/\kf, \sigma/ \kappa \kf^2)
    \nonumber\\
\tilde{\gamma}^{(B)}_z &= u_{\rm rms} \, \Rm \; g (s, \qeta, k/\kf, \sigma/ \eta \kf^2)
\label{eq289}\\
g (s, q, v, w) &=  \frac{q}{2} \left( \phantom{+}\frac{1 + s^2 + v}{\big((1 + v)^2 + s^2 + w\big)^2 + q^2}\right.
\nonumber\\
&\quad\quad\;\:
    + \left. \delta \frac{1 + s^2 - v}{\big((1 - v)^2 + s^2 + w\big)^2 + q^2} \right)
\nonumber
\end{align}
and
\begin{alignat}{2}
\tilde{\kappa}_{zz} &=&& - \kappa_{{\rm t}0} \, \Pe \;\: h (s, \qkap, k/k_{\rm f}, \sigma/ \kappa \kf^2)
\nonumber\\
\tilde{\eta}_{xx} &= \tilde{\eta}_{yy} = &&- \eta_{{\rm t}0} \, \Rm \, h (s, \qeta, k/\kf, \sigma/ \eta \kf^2)
\label{eq291}
\end{alignat}
\begin{align}
h (s, q, v, w) &= - \frac{1}{2 v} \left( \phantom{-}\frac{(1 + s^2 + v) \big((1 + v)^2 + s^2 + w\big)}{\big((1 + v)^2 + s^2 + w\big)^2
    + q^2}\right.
\nonumber\\
\qquad
    &\qquad\quad\;\;-\left. \delta \frac{(1 + s^2 - v) \big((1 - v)^2 + s^2 + w\big)}{\big((1 - v)^2 + s^2 + w\big)^2 + q^2} \right).
\nonumber
\end{align}
The factor $\delta$ is in general equal unity but equal to zero if $1 - v = s = w = q = 0$,
that is, if the following denominator vanishes.
All coefficients $\gamma^{(C)}_i\!\!$, $\gamma^{(B)}_i\!\!$,
$\tilde{\gamma}^{(C)}_i\!\!$, $\tilde{\gamma}^{(B)}_i\!\!$, $\kappa_{ij}$, $\eta_{ij}$, $\tilde{\kappa}_{ij}$,
and $\tilde{\eta}_{ij}$, which are not explicitly mentioned, are equal to zero.

\begin{figure}[t!]\begin{center}
\includegraphics[width=\columnwidth]{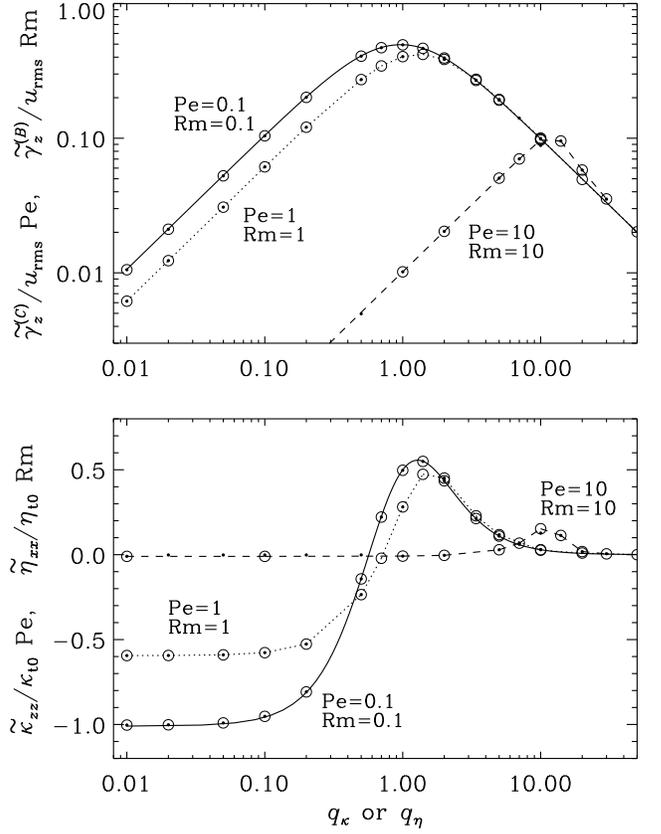}
\end{center}\caption[]{
Dependence of $\tilde{\gamma}^{(C)}_{z} / u_{\rm rms} \Pe$ and $\tilde{\kappa}_{zz} / \kappa_{{\rm t}0} \Pe$
on $\qkap$
as well as that of $\tilde{\gamma}^{(B)}_z / u_{\rm rms} \Rm$
and $\tilde{\eta}_{xx} / \eta_{{\rm t}0} \Rm$ on $\qeta$
for the model given by Eq.~\eqref{eq272} for $k/\kf = 0.1$,  $s=0.01$,
$\sigma =0$ and three values of $\Pe$ or $\Rm$, respectively.
Solid lines give the SOCA results,
symbols the values obtained by the test-field method;
dots correspond to $\tilde{\gamma}^{(C)}_{z} / u_{\rm rms} \Pe$ and $\tilde{\kappa}_{zz} / \kappa_{{\rm t}0} \Pe$,
open circles to $\tilde{\gamma}^{(B)}_z / u_{\rm rms} \Rm$,
and $\tilde{\eta}_{xx} / \eta_{{\rm t}0} \Rm$.
Clearly, $\tilde{\gamma}^{(C)}_{z} / u_{\rm rms} \Pe$ and $\tilde{\kappa}_{zz} / \kappa_{{\rm t}0} \Pe$
coincide completely with $\tilde{\gamma}^{(B)}_z / u_{\rm rms} \Rm$ and $\tilde{\eta}_{xx} / \eta_{{\rm t}0} \Rm$
if $\Pe$ and $\Rm$ coincide.
}\label{pwave_kap}\end{figure}

Numerical calculations of $\tilde{\gamma}^{(C)}_z$ and $\tilde{\kappa}_{zz}$
as well as $\tilde{\gamma}^{(B)}_z$ and $\tilde{\eta}_{xx}=\tilde{\eta}_{yy}$
by the test-field method, without restriction to SOCA, have been carried out
with $k / \kf = 0.1$ and some specific values of $\Pe$ and $\Rm$.
Only cases with  $s \ne 0$ were included, for which the $\chi$ and $x$
averages are equivalent.
Hence the standard test-field procedure with horizontal averages could be employed.
Figure~\ref{pwave_kap} shows these quantities for $s = 0.01$
as functions of $\qkap$ or $\qeta$.
The results for $\Pe = 0.1$ and $\Rm = 0.1$ are in good agreement with our SOCA calculations,
that is, \eq{eq289} and \eq{eq291}.
Those for higher $\Pe$ and $\Rm$ clearly deviate from them.
Interestingly, the dependence of $\tilde{\gamma}^{(C)}_z$ and $\tilde{\kappa}_{zz}$ on $\Pe$ is
always the same as those of $\tilde{\gamma}^{(B)}_z$ and $\tilde{\eta}_{xx}$ or $\tilde{\eta}_{yy}$ on $\Rm$.

A remarkable feature of the SOCA results \eq{eq291} for
$\tilde{\kappa}_{zz}$, and also for $\tilde{\eta}_{xx} = \tilde{\eta}_{yy}$,
is that these quantities show singularities at $k / \kf = 1$ if $\qkap = \qeta = s = \sigma = 0$.
Nevertheless they are well defined at this point; $\tilde{\kappa}_{zz} / \kappa = \Pe^2 / 4$
and $\tilde{\eta}_{xx} / \eta =\tilde{\eta}_{yy} /\eta = \Rm^2 / 4$ at $k / \kf = 1$.

\begin{figure}[t!]\begin{center}
\includegraphics[width=\columnwidth]{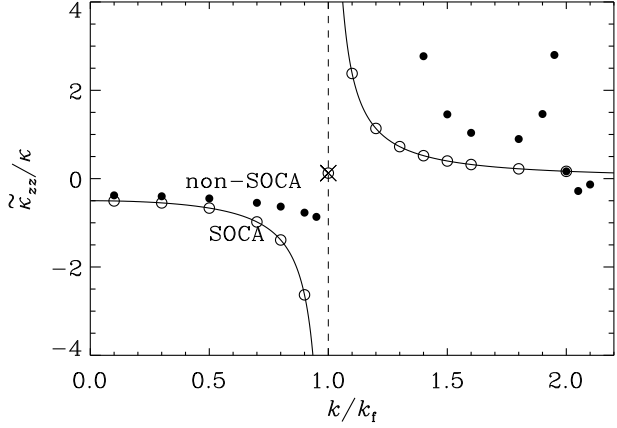}
\end{center}\caption[]{
Dependence of $\tilde{\kappa}_{zz} / \kappa$ with $\Pe=0.707$
and $q_\kappa = \sigma = 0$ on $k/\kf$.
Solid lines as well as the cross at $k/\kf = 1$
result from analytic SOCA calculations with $s=0$.
Symbols give numerical results obtained with Eq.~\eq{eq27} using $s=0.01$;
filled circles: full equation,
open circles: SOCA, term $(\uu c)'$ dropped.
Note the second ``resonance" at $k/\kf=2$.
}\label{pC1_n128_kdep}\end{figure}

Let us, in what follows, focus attention on passive scalars only.
Consider a mean scalar of the form $\meanC = F(t) \cos kz$ with $F$ being positive.
For $q_\kappa = 0$ its time behavior is exclusively determined by the quantity $\kappa + \tilde{\kappa}_{zz}$.
Clearly $\meanC$ is bound to decay if $\kappa + \tilde{\kappa}_{zz} > 0$.
Now consider the dependence of $\tilde{\kappa}_{zz}$ on $k / \kf$ for $\qkap = s = \sigma = 0$
depicted in Fig.~\ref{pC1_n128_kdep}.
If $k/\kf$ is smaller than but close to unity, $\tilde{\kappa}_{zz}$ may, even for small $\Pe$,
take arbitrarily large negative values, and $\kappa + \tilde{\kappa}_{zz}$ becomes negative.
This will then lead to a growth of the modulus of $\meanC$.
Of course, this conclusion is drawn from a result obtained under the SOCA
and may hence be questionable.
Indeed, the sufficient condition for the applicability of SOCA given so far, $\Pe \ll 1$,
has been derived for $k/\kf \ll 1$ only.
If, by contrast, $k/ \kf \approx 1$, we find, when comparing the terms $\nabla\cdot(\uu c)'$ and $\kappa\Delta c$
in \eq{eq27} under the assumption that $c$ is dominated by contributions with wavenumbers $k+\kf$ and $k-\kf$,
for $q_\kappa = \sigma=0$ and $s\ll 1$ the more stringent condition
\EQ
\Pe \ll 3 \frac{(1-k/\kf)^2+s^2}{1-k/\kf + s^2} \, .
\label{eq301}
\EN
It supports the doubts in the above conclusion concerning the growth of the modulus of $\meanC$.

The aforementioned SOCA calculations for $s = 0$ have been extended by the inclusion of fourth-order terms in $\uu$,
that is, in $\Pe$.
Apart from some quantitative changes of $\tilde{\kappa}_{zz} / \kappa$
in the neighborhood of $k/\kf = 1$,
which occur with larger $\Pe$, a new singularity emerges at $k/\kf = 2$.
As can be seen in Fig.~\ref{pC1_n128_kdep}, the numerical (non-SOCA)
calculations with $s = 0.01$ reflect this feature, too.
They also give indications of a further resonance at $k/\kf = 3$ (not shown).

\begin{figure}[t!]\begin{center}
\includegraphics[width=\columnwidth]{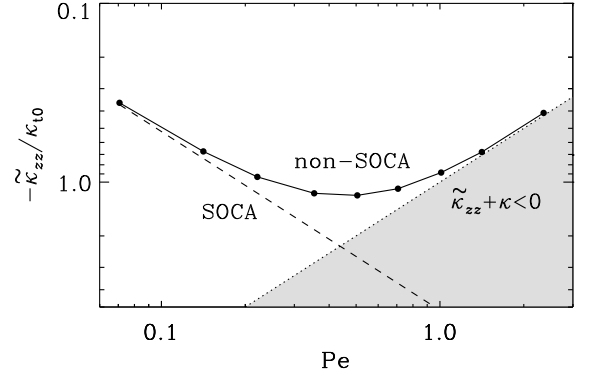}
\end{center}\caption[]{
Dependence of $\tilde{\kappa}_{zz} / \kappa_{{\rm t}0}$ on $\Pe$ for $k/\kf = 0.9$ and $s=0.01$
$\sigma = q_\kappa = 0$.
The dashed line gives the SOCA result and the shaded area marks the range where the total diffusivity would become negative.
}\label{ppot2_kappa}\end{figure}

Figure~\ref{ppot2_kappa} shows $\tilde{\kappa}_{zz}$ for steady test
fields (that is, $\sigma = 0$) with $k/\kf = 0.9$ and $s = 0.01$
as a function of $\Pe$.
The results clearly deviate from those obtained by the SOCA as soon as $\Pe$ exceeds, say, $0.2$.
Considering that $\kappa + \tilde{\kappa}_{zz} \geq 0$ is equivalent to $- \tilde{\kappa}_{zz} / \kappa_{\rm t0} \leq 1 / \Pe$,
Figure~\ref{ppot2_kappa} tells us further that $\kappa + \tilde{\kappa}_{zz}$ becomes very small with growing $\Pe$,
but suggests that it remains positive.
We may suppose that the modulus of the considered $\meanC$
never grows but its decay becomes very slow for large $\Pe$.
For example, for $\Pe=1$ we expect that $\lambda = - (\kappa + \tilde{\kappa}_{zz}) k^2  \approx - 0.1 \kappa k^2$,
that is, the decay of $\meanC$ should be about ten times slower than in the absence of any motion.

\begin{figure}[t!]\begin{center}
\includegraphics[width=\columnwidth]{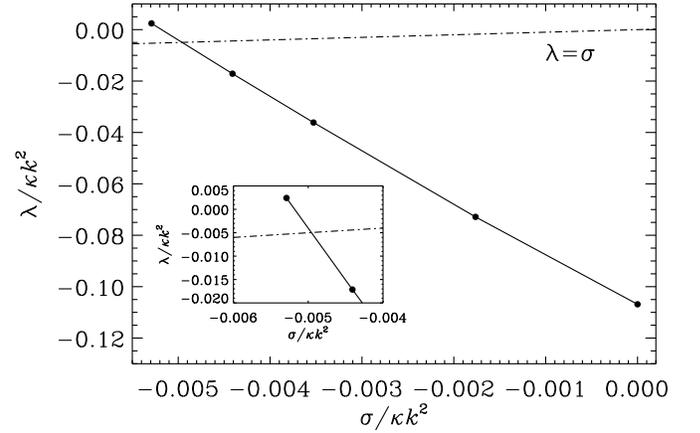}
\end{center}\caption[]{
Dependence $\lambda(\sigma) = - [(\kappa+\tilde{\kappa}_{zz}(k, \sigma)]k^2$
for $\Pe=1.0$, $k/\kf=9/10$, $s=0.01$, and $q_\kappa=0$.
The curve representing $\lambda (\sigma) / \kappa k^2$
intersects the dash-dotted line $\lambda = \sigma$
at $\lambda (\sigma) / \kappa k^2 = -0.005$,
which is shown more clearly in the inset.
}\label{ppot2_kappa_lam}\end{figure}

In these considerations, however, the memory effect,
that is, the dependence of the value of $\tilde{\kappa}_{zz}$, relevant for the decay of $\meanC$, on the decay rate $\lambda$
itself, has been ignored.
As explained in Sec.~\ref{subsIIIB}, we
have to include this dependence by using time-dependent test fields.
Let us assume that they vary as $\exp (\sigma t)$ but consider $\sigma$ first as independent of $\lambda$.
Then $\tilde{\kappa}_{zz}$ and $\lambda = - (\kappa + \tilde{\kappa}_{zz}) k^2$ occur as functions of $\sigma$.
Figure~\ref{ppot2_kappa_lam}, obtained by test-field calculations, shows this dependence of $\lambda$ on $\sigma$.
If we then identify $\sigma$ with $\lambda$ we find, as indicated in Fig.~\ref{ppot2_kappa_lam},
$\lambda \approx-0.005\kappa k^2$.
That is, the decay of $\meanC$ is about 200 times slower than in the absence of any motion.

\begin{figure}[t!]\begin{center}
\includegraphics[width=\columnwidth]{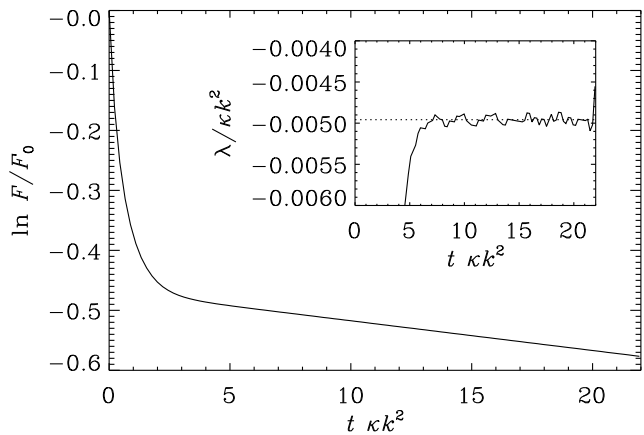}
\end{center}\caption[]{
Direct simulation showing the decay of
the amplitude $F$ of $\meanC(z)$
in units of its initial value $F_0$.
Parameters like in Fig.~\ref{ppot2_kappa_lam}.
The inset shows the time dependence of
the growth rate $\lambda$, leveling off
at $\lambda\approx-0.005\,\kappa k^2$
after some initial adjustment time.
}\label{plam}\end{figure}

\begin{figure}[t!]\begin{center}
\includegraphics[width=\columnwidth]{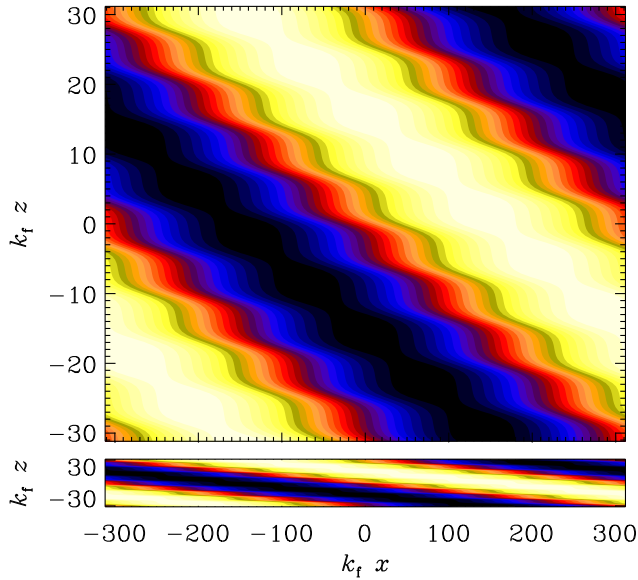}
\end{center}\caption[]{
(Color online) Snapshot of $C(x,z)$ for the simulation shown in \Fig{plam}
at $t\kappa k^2=17$.
Bright (yellow) shades indicate positive values and dark (blue) shades
negative values.
The lower panel shows $C(x,z)$ with the correct aspect ratio of the box.
}\label{ppvar}\end{figure}

In order to check this surprising result, we perform two-dimensional
direct numerical simulations based on Eq. \eq{eq21}
with a flow given by \eq{eq272} using $k_0=10\,k_1$, where $k_1=2\pi/L_z$
is the smallest wave number in the $z$ direction with extent $L_z$.
Our computational domain is periodic in both directions.
We choose $L_x=10\,L_z$ so as to accommodate the variation in the
$x$ direction with wavenumber $k_0 s$ and $s=0.01$.
The initial condition is $C=C_0\cos kz$,
with $C_0>0$ and $k=9\,k_1$.
We use $128^2$ mesh points and choose $\Pe=1$, which is clearly
beyond the applicability of SOCA; cf. Fig.~\ref{ppot2_kappa}.

As we expect that $\meanC = F(t) \cos kz$,
we have identified the maximum of the $x$-average of $C$
with respect to $z$ with $F$ and determined
the growth rate by calculating first
its instantaneous value $\lambda(t) = \dd\ln F/\dd t$; see Fig.~\ref{plam}.
It turns out that the average
of $\lambda$ over the time interval, in which it
is approximately constant
is in excellent agreement with the test-field result $\lambda\approx-0.005\kappa k^2$, described above.
The snapshot in Fig. \ref{ppvar} shows that $C(x,z)$ varies mainly in
the $z$ direction with the dominant wave number $k=9 \kf/10$.
In units of $\kappa k^2$, the
{\em free-decay rate} of a mode with this wave number would be
$0.01$, or $0.012$, if the variation in $x$ is taken into account.
Note, however, that the
dominant constituent of $C$ belongs, by virtue of its $x$ dependence,
to the {\em fluctuating} field $c$ and that the
actual decay rate of the {\em mean field} $\meanC$
is at least two times smaller than the given free-decay rate.
The fluctuations are not decaying freely, but
follow the mean field, and hence adopt its decay rate.
In this particular case, the rms values of the fluctuations
exceed those of the mean field by a factor of 14.

The question could be raised as to whether a resonance effect in the above sense
can also occur for solenoidal flows.
Numerical experiments with the (stationary) ABC flow
(for its definition see, e.g., \citep{Dombre}) indicate
clearly that the decay of $\meanC$ is always accelerated in the presence
of this flow, irrespective of the value of $k/\kf$.

\section{Conclusions}

In this work we have shown that the turbulent diffusivity $\kappat$ for the
concentration of a passive scalar in a potential flow can be negative
at low P\'eclet numbers.
This result is analogous to an earlier finding for the turbulent magnetic
diffusivity $\etat$ in such a flow at low magnetic Reynolds numbers,
originally derived in the context of astrophysical dynamo theory.
The numerical calculations presented in this paper confirm \Eq{eq01} quantitatively for an irrotational flow.
We have not yet considered
the case of the combined action of solenoidal and irrotational flows
where the question arises of how strong the solenoidal part
has to be to render the turbulent diffusivities positive.
Our calculations also show that negative values of $\kappat$ do not occur for larger P\'eclet numbers,
whereas negative $\etat$ may well exist for moderate to large Reynolds numbers.
In neither case have negative turbulent diffusivities
yet been seen in laboratory experiments.
Nevertheless, for possible physical applications of our results
one may think of microfluidic devices \cite{microfluidic},
in which the flow can be compressible \cite{microfluidic_comp}
and the P\'eclet number small.

In addition to the condition of small P\'eclet and magnetic
Reynolds numbers, there are also the requirements of
good scale separation and of slow temporal variations of the flow.
If these requirements are not obeyed,
$\kappat$ and $\etat$ are no longer necessarily negative --
even at small values of P\'eclet and magnetic Reynolds numbers.
This may be the reason why a reduction of the effective diffusivity
has never been seen in physically meaningful compressible flows
and why \Eq{eq01} is virtually unknown in the turbulence community.
In fact, previous attempts to verify this equation in simulations
have failed because of the fact that the time dependence
has been too vigorous in those flows \cite{BDS10}.

The spatial structure of the flow does not appear to be critical
for obtaining a reduction of the effective diffusivity.
Even in a nearly one-dimensional flow, turbulent diffusivities can become negative.
However, in that case there are two new effects.
First, if the underlying flow pattern displays propagating wave
motions, there can be transport of the mean scalar in the direction of wave propagation
-- even in the absence of any mean material motion.
Again, this effect may have applications to microfluidic devices.
Second, the wavenumber characteristics display a singular behavior
under SOCA, but even beyond SOCA there can be a dramatic slow-down
of the decay by factors of several hundreds compared with the molecular values.
This result is completely unexpected because no such behavior has
ever been seen for any other turbulent transport process.
Furthermore, the memory effect proves to be markedly important in
such cases, so the common assumption of
an instantaneous relation between the mean flux of the scalar and its mean concentration or
the mean electromotive force and the mean magnetic field
breaks down near the singularity.

In addition to finding out more about possible applications of the
turbulent transport phenomena discussed above, it would be natural to
study the possibility of similar processes for the turbulent transport
of other quantities including momentum and heat or other active scalars.
Further, a complementary effort to determine the transport coefficients for
{\em turbulent} irrotational flows numerically would be of high interest,
the more as there are no simple analytical results available.
Supernova-driven turbulence in the interstellar medium would
of course be the most suggestive application.

Clearly, both analytical and numerical approaches using the test-field
method proved to be invaluable in that they
are able to predict unexpected phenomena that can then also be verified using direct
numerical simulations and in future hopefully also laboratory experiments.

\appendix

\section{Relations for $a_{ij}$, $\eta_{ij}$, and $c_{ijk}$}
\label{appA}

Under SOCA we may derive, for homogeneous turbulence,
\EQ
a_{ij} = \int \!\!\! \int \ii (\epsilon_{ilm} k_j -\epsilon_{ilj} k_m) \frac{ \hat{Q}_{lm} (\kk, \omega)
    }{\eta k^2 - \ii \omega} \, \dd^3k \, \dd \omega \, ,
\label{eqA01}
\EN
\EQA
\eta_{ij} \!\!&=&\!\!\frac{1}{2}\!\int \!\!\! \int \left(\delta_{ij} \delta_{lm} - \delta_{im} \delta_{jl}
   - \frac{2 \eta (\delta_{ij} k_l k_m - k_i k_m \delta_{jl})}{\eta k^2 - \ii \omega} \right)
\nonumber\\
&& \qquad \qquad \qquad
    \times \frac{\hat{Q}_{lm} (\kk, \omega)}{\eta k^2 - \ii \omega}\, \dd^3k \, \dd \omega \, ,
\label{eqA03}
\ENA
\EQA
c_{ijk} \!\!&=&\!\! - \frac{1}{2} \int \!\!\! \int
    \bigg(2 \epsilon_{imn} \delta_{jk}
    - (\epsilon_{imj} \delta_{kn} + \epsilon_{imk} \delta_{jn})
\nonumber\\
&& \qquad
    - 2 \eta\! \left.\frac{ 2 \epsilon_{imn} k_j k_k - (\epsilon_{imj} k_k + \epsilon_{imk} k_j) k_n }{\eta k^2 - \ii \omega}\!\!\right)\!
\nonumber\\
&& \qquad \qquad \qquad \times  \frac{\hat{Q}_{mn} (\kk, \omega)}{\eta k^2 - \ii \omega} \, \dd^3k \, \dd \omega \, .
\label{eqA05}
\ENA

\acknowledgments
We thank the anonymous referee for fruitful hints that helped to improve the paper.
K.-H.R.\ and A.B.\ are grateful for the opportunity to work on this paper
while participating in the program ``The Nature of Turbulence"
at the Kavli Institute for Theoretical Physics in Santa Barbara, CA.
This work was supported in part by
the European Research Council under the AstroDyn Research Project No.\ 227952,
the Swedish Research Council Grant No.\ 621-2007-4064,
and the National Science Foundation under Grant No.\ NSF PHY05-51164.
We acknowledge the allocation of computing resources provided by the
Swedish National Allocations Committee at the Center for
Parallel Computers at the Royal Institute of Technology in
Stockholm and the National Supercomputer Centers in Link\"oping.



\begin{thebibliography}{}

\bibitem{EKR95}
T. Elperin, N. Kleeorin, and I. Rogachevskii\yprl{1995}{52}{2617}

\bibitem{EKR96}
T. Elperin, N. Kleeorin, and I. Rogachevskii\yprl{1996}{76}{224}

\bibitem{EKR97}
T. Elperin, N. Kleeorin, and I. Rogachevskii\ypre{1997}{55}{2713}

\bibitem{BSV09}
A. Brandenburg, A. Svedin, and G. M. Vasil\ymn{2009}{395}{1599}

\bibitem{HKRB11}
N. E. L. Haugen, N. Kleeorin, I. Rogachevskii, and A. Brandenburg\sprl{2011},
arXiv:1101.4188

\bibitem{Mof78}
H. K. Moffatt\ybook{1978}
{\em Magnetic field generation in electrically conducting fluids}
{Cambridge University Press, Cambridge}

\bibitem{KR80}
F. Krause and K.-H. R\"adler\ybook{1980}
{\em Mean-field magneto\-hydro\-dy\-na\-mics and dynamo theory}
{Pergamon Press, Oxford}

\bibitem{Rae00}
K.-H. R\"adler, in {\em From the Sun to the Great Attractor}
edited by D. Page and J. G. Hirsch,
Lecture Notes in Physics Vol. 556 (Springer, Berlin, 2000), p. 101.

\bibitem{Rae07}
K.-H. R\"adler, in {\em Magnetohydrodynamics: Historical Evolution and Trends},
edited by S. Molokov, R. Moreau and H. K. Moffatt
(Springer, Dordrecht, 2007), p. 55.

\bibitem{RR07}
K.-H. R\"adler and M. Rheinhardt\ygafd{2007}{101}{11}

\bibitem{Korpi_etal99}
M. J. Korpi, A. Brandenburg, A. Shukurov, I. Tuominen, and \AA. Nordlund\yapjl{1999}{514}{L99}

\bibitem{dAML02}
M. A. de Avillez and M.-M. Mac Low\yapj{2002}{581}{1047}

\bibitem{Bals04}
D. S. Balsara, J. Kim, M.-M. Mac Low, and G. J. Mathews\yapj{2004}{617}{339}

\newpage

\bibitem{Gre08}
O. Gressel, D. Elstner, U. Ziegler, and G. R\"udiger \yana{2008}{486}{L35}

\bibitem{DSB11}
F. Del Sordo and A. Brandenburg\yana{2011}{528}{A145}

\bibitem{KKS86}
K. Kajantie and H. Kurki-Suonio\yprd{1986}{34}{1719}

\bibitem{Ingatius94}
J. Ignatius, K. Kajantie, H. Kurki-Suonio, and M. Laine\yprd{1994}{49}{3854}

\bibitem{LL}
E. M. Lifshitz and L. P. Pitaevskii\ybook{1981}
{Physical kinetics}
{1st ed., Pergamon Press, Oxford}

\bibitem{Chatterjee}
P. Chatterjee, D. Mitra, M. Rheinhardt, and A. Brandenburg\pana{2011}
arXiv:1011.1218,  DOI:10.1051/0004-6361/201016108.

\bibitem{HB09}
A. Hubbard and A. Brandenburg\yapj{2009}{706}{712}

\bibitem{SRSRC07}
M. Schrinner, K.-H. R\"adler, D. Schmitt, M. Rheinhardt, and U. R. Christensen\ygafd{2007}{101}{81}

\bibitem{BRS08}
A. Brandenburg, K.-H. R\"adler, and M. Schrinner\yana{2008}{482}{739}

\bibitem{PC}
The Pencil Code is a high-order finite-difference code
(sixth order in space and third order in time);
\url{http://pencil-code.googlecode.com}.

\bibitem{Dombre}
T. Dombre, U. Frisch, J. M. Greene, M. H\'enon, A. Mehr, and A. M. Soward\yjfm{1986}{167}{353}

\bibitem{microfluidic}
J. Koo and C. Kleinstreuer\yjour{2003}{J.\ Micromech.\ Microeng.}{13}{568}

\bibitem{microfluidic_comp}
Z. Yao, P. Hao, and X. Zhang\yjour{2011}{Sci.\ China Phys., Mech.\ Astron.}{54}{711}

\bibitem{BDS10}
Brandenburg, A., \& Del Sordo, F.\yproc{2010}{432}
{Highlights of Astronomy, Vol. {\bf 15}}
{E. de Gouveia Dal Pino}
{CUP}

\end{thebibliography}
\end{document}